\documentclass{elsart}[12pt]
\usepackage[dvips]{graphicx}
\begin{document}
\begin{frontmatter} 
\title {Charge and current-sensitive preamplifiers  \\
for pulse shape discrimination techniques with silicon detectors} 
\small
\author[IPNO]{H. Hamrita},
\author[IPNO]{E. Rauly},
\author[IPNO]{Y. Blumenfeld},  
\author[IPNO]{B. Borderie\thanksref{cores}},
\author[IPNO]{M. Chabot}, 
\author[IPNO]{P. Edelbruck},
\author[IPNO]{L. Lavergne},
\author[IPNO]{J. Le Bris},
\author[LPC]{Th. Legou},
\author[IPNO]{N. Le Neindre},
\author[IPNO]{A. Richard},
\author[IPNO]{M.F. Rivet},
\author[IPNO]{J. A. Scarpaci},
\author[LPC]{J. Tillier}
\author[IPNO]{S. Barbey},  
\author[IPNO]{E. Becheva},  
\author[LPC]{F. Bocage},
\author[LPC]{R. Bougault},
\author[IPNO]{R. Bzyl},
\author[LPC]{S. Gr\'evy},
\author[LPC]{B. Carniol},
\author[LPC]{D. Cussol}, 
\author[IPNO]{P. D\'esesquelles},
\author[LPC]{D. Etasse},
\author[IPNO,AEM]{E. Galichet},
\author[LPC]{S. Gr\'evy},
\author[IPNL]{D. Guinet},
\author[IPNO]{G. Lalu},
\author[CATA1]{G. Lanzalone},
\author[IPNL]{Ph. Lautesse},
\author[LPC]{O. Lopez},
\author[IPNO]{G. Martinet},
\author[IPNO]{S. Pierre},
\author[CATA2]{G. Politi},
\author[NAP]{E. Rosato}, 
\author[LPC]{B. Tamain}
\and
\author[LPC]{E. Vient}

\address[IPNO]{Institut de Physique Nucl\'eaire, IN2P3-CNRS, F-91406
Orsay Cedex, France} 
\address[LPC]{LPC, IN2P3-CNRS, ENSICAEN et Universit\'e,
F-14050 Caen  Cedex, France}
\address[AEM]{Conservatoire National des Arts et M\'etier,
F-75141 Paris Cedex 03, France}
\address[IPNL]{Institut de Physique Nucl\'eaire, IN2P3-CNRS et 
Universit\'e, F-69622 Villeurbanne Cedex, France}
\address[CATA1]{INFN, Laboratori Nazionali del Sud and Dipartimento di
  Fisica e Astronomia, Universit\`a di Catania, Italy}
\address[CATA2]{INFN, Sezione di Catania and Dipartimento di
  Fisica e Astronomia, Universit\`a di Catania, Italy}
\address[NAP]{Dipartimento di Scienze Fisiche e Sezione INFN, Univ. di Napoli
 ``Federico II'', I80126 Napoli, Italy} 
\thanks[cores]{Corresponding author. Tel 33 1 69157148; fax 33 1
69154507; \\e-mail borderie@ ipno.in2p3.fr} 
\normalsize
\begin{abstract} 
New  charge and current-sensitive preamplifiers coupled to silicon
detectors and devoted to studies in nuclear structure
and dynamics have been developed and tested. For the first time shapes of
current pulses from light charged particles and carbon ions are presented.
Capabilities for pulse shape discrimination techniques are demonstrated.
\end{abstract} 
\end{frontmatter}
\section{Introduction} 
In a near future, beams of nuclei farther and farther away from stability
should become available. This will enlarge the field of nuclear physics studies
related to the influence of the N/Z degree of freedom. Investigating the
effects
of this parameter will however require the identification of both charge and
mass numbers of nuclei produced in nuclear collisions, over the largest
possible
range. To attain this goal pulse-shape
discrimination (PSD) of signals from solid state detectors appears as the
most promising avenue.
This method proposed more than forty years ago\cite{AM63} was recently
 investigated in detail in relation with the design of a 4$\pi$ silicon 
ball\cite{PA94,PA96,PA97,PA00}. Moreover potentially improved capabilities
have been also discussed using homogeneously doped n-TD silicon
detector~\cite{MU00,VON92} or PSD-time of flight coupling~\cite{LU01}.
In all the recent studies PSD was obtained by using conventional
charge-sensitive preamplifiers and by injecting particles or ions into
the rear-side (n-side) of totally depleted detectors. This injection mode,
first proposed in ref.~\cite{AM63}, presents two major advantages: a rising
field profile with penetration depth (the highest ionization density is
associated to the highest field) and a monotonic increase of charge collection
time with charge and mass number of the detected particle or ion, which
facilitates identification.

For the study of nuclear structure through direct reactions double sided
silicon-strip detectors, with narrow pitch ($\leq$ 1mm), are generally used
to measure the energy and position of recoiling light charged
particles~\cite{BL99}. For studies in nuclear dynamics, resolution in 
position is less crucial, but Z and A identification is needed for both
particles and fragments~\cite{PO95}. 

In this work, charge and current-sensitive preamplifier prototypes for
 nuclear structure and dynamics experiments have been
developed and tested
with the aim of improving PSD method by studying in detail current signal
shapes from particles and ions over a large energy range. Note that current
signal shapes have been recently used in atomic cluster studies to identify
partitions of carbon cluster fragmentation~\cite{CHA02}.

The paper is organized as follows. In section 2 the general guidelines
of the design of preamplifiers are introduced. Section 3 is devoted to 
implementation and characterization of preamplifiers. In section 4 results of 
beam tests will be presented, discussed and compared to a simple
simulation.
\section{Preamplifier design guidelines} 
The requirements which have been considered in the design of preamplifiers
are the following:

(1) Preamplifiers must be mounted in vacuum, as close as possible to their
corresponding silicon detector to minimize the series inductance from the
connection lead. Preamplifiers must have small size and low power
consumption.

(2) Keeping the conventional part (charge-sensitive) to measure energies, an
additional circuit delivering an accurate image of the current signal
(converted into
a voltage) is required. Such a signal will permit deeper investigations of
PSD capabilities as compared to present risetime signal analysis using
charge-sensitive preamplifier.

(3) For structure studies a conversion gain of $\sim$ 15 mV/MeV with an
energy
range up to 120 MeV is well suited for the detection of particles and light
ions (Z$\leq 10$). For dynamics studies a lower conversion gain is necessary
to cover the large energy range needed $\sim$ 3 GeV for particles and ions.

(4) For structure studies silicon detectors have typical capacitances
in the range 10-35 pF whereas for dynamics studies detectors with
capacitances in the range 60-200 (600) pF are required. Preamplifiers (with
possibly different versions) must match different detectors.

(5) Energy resolution (preamplifier + amplifier + analog to
digital conversion) for structure studies (dynamics studies) better than
50 keV for 5 MeV protons  (0.5\% for 80 MeV carbon ions) is demanded.
\section{Current and charge-sensitive preamplifiers}  
A new type of preamplifier with two outputs has been designed. It
will provide simultaneously :
\begin{itemize}
\item [$\bullet$] an accurate measurement of the overall electrical
  charge collected from the
detector (as for classical spectroscopy) : ``charge output''

\item [$\bullet$] a high bandwidth, precise representation of the detector
collection current over time : ``current output''
\end{itemize}
\subsection{Principle of operation}
\subsubsection{General}
\begin{figure}[htbp]
\centerline{\includegraphics*[scale=0.8]{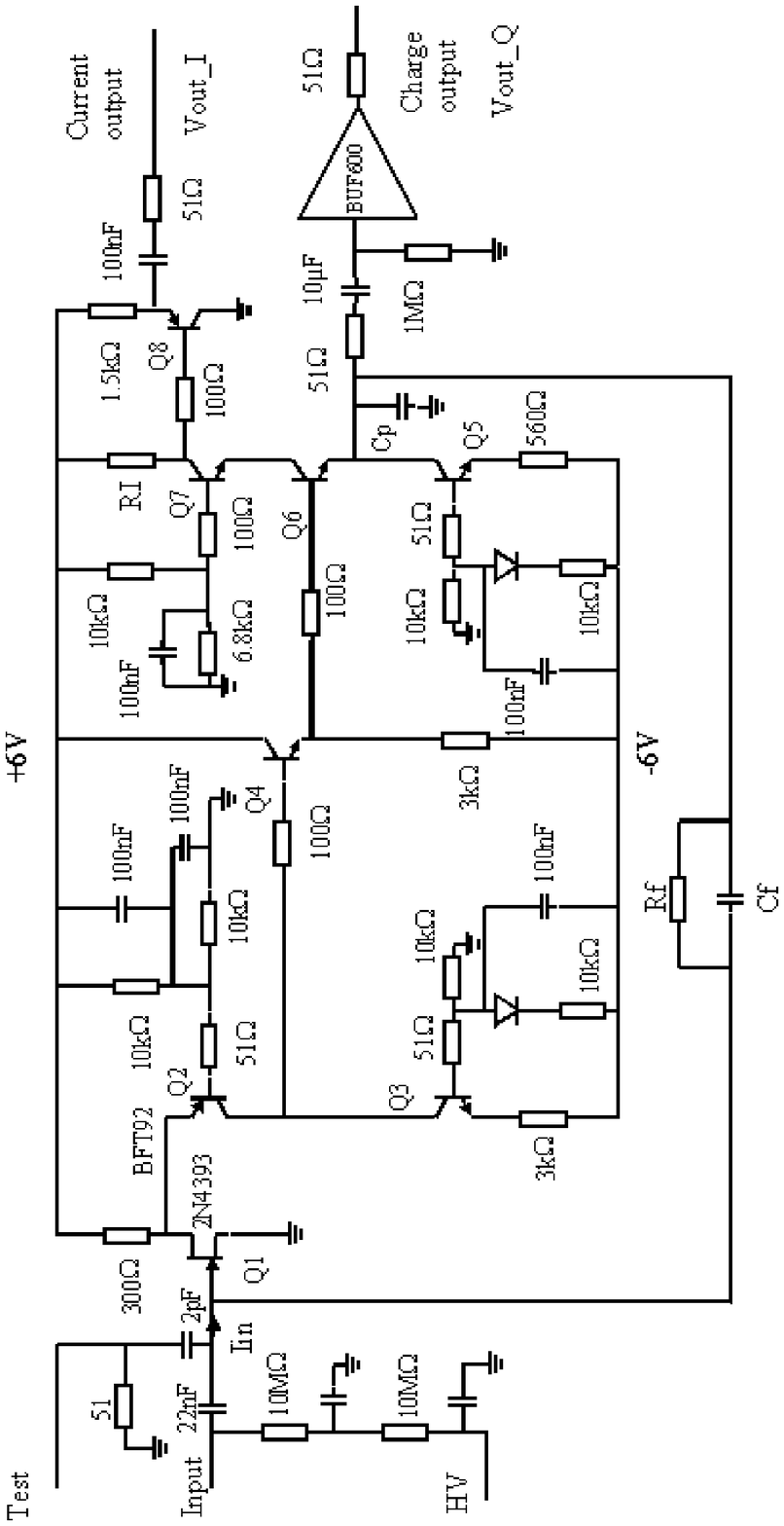}}
\caption{Diagram of the dual output preamplifier (high gain version).}
\label{Fig1}
\end{figure}

\begin{figure}[htbp]
\centerline{\includegraphics*[scale=0.8]{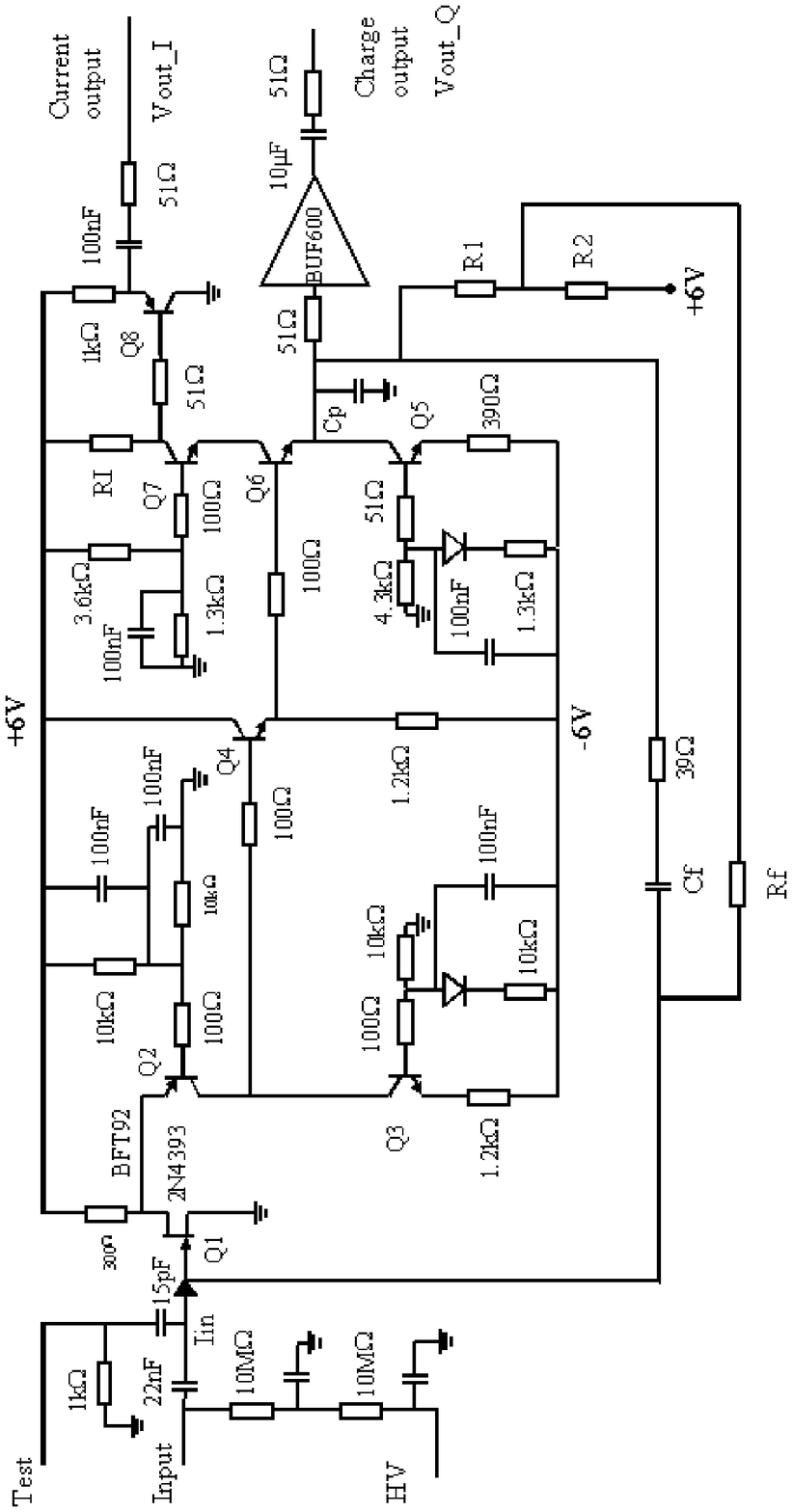}}
\caption{Diagram of the dual output preamplifier (low gain version).}
\label{Fig2}
\end{figure}

The device is based on a classical folded cascode charge sensitive
preamplifier (CSA) (see figures~\ref{Fig1} and~\ref{Fig2}). The first stage
uses a high frequency Field
Effect Transistor (Q1) associated to a common base PNP
transistor (Q2). The FET has been chosen for its high
transconductance (g$_{m})$ associated to a reduced cut-off voltage
dispersion V$_{gs(off)}$ . The g$_{m }$ can reach 25 mS for V$_{gs(off)}$ =
-3V.
Q2 is fed by a current source (Q3). An emitter follower (Q4)
drives the signal to the output stage (Q6-Q7). The feedback
capacitor Cf and resistor Rf of the CSA are connected between the
emitter of Q6 and the input, while the charge output is driven to the
external world through an integrated buffer (BUF600).

The output stage is composed of an emitter follower Q6 fed by a current
source (Q5). The current flowing through the feedback network is
provided by the emitter of Q6.
A common base stage (Q7) has been inserted in
the collector path of Q6 in order to convert the output current into a
voltage image which is driven out through the emitter follower stage Q8.
This voltage represents the ``current output'' of the amplifier
(Vout\_I$\sim$Ie(Q7) x RI and Ie(Q6)$\sim$Ic(Q6)$\sim$Ie(Q7)$\sim$Ic(Q7)).
Actually, the input impedance of Q1 being very high, this current
corresponds to the detector current. The current gain of the preamplifier
(Vout\_I/Iin) is proportional to the value of RI. However, several other
factors affect the actual gain value, the major one being the additional
capacitance load present on the emitter of Q7 (stray capacitance + input
capacitance of the integrated buffer). This increases the gain by a factor
1+Cp/Cf. Cp has been experimentally determined to be 5 pF. The output stage
Q8 also provides an attenuation which can not be neglected. The gain is
eventually determined by simulations including the stray capacitances in each 
preamplifier version. Note that both outputs are matched to 50 $\Omega$
with a serial resistor.

\subsubsection{Simulated high gain version - structure studies}
For the charge output, the gain (open output) is equal to 13 mV/MeV with
Cf = 3.3 pF, which permits energy measurements up to 120 MeV corresponding to
an output swing Vout\_Q of 1.5 V. The current gain (transimpedance) has been
determined by simulation to 550 $\Omega$
\subsubsection{Simulated low gain version - dynamics studies}
A higher energy dynamic range is needed. This has been achieved by 
reducing the preamplifier gain (higher Cf) and moving the output operating
point towards the negative supply in order to increase the possible voltage
swing. This offset has been introduced thanks to the voltage divider R1/R2
(274 k$\Omega$/620 k$\Omega$) referenced to the +6 V supply voltage (rest
output voltage equal to -3.7 V). The bias voltages of Q5 and Q7 have been
accordingly adjusted. The charge output gain is equal to 1.47 mV/MeV with 
Cf = 30 pF, leading to an input swing voltage of 3.7 V for an energy of 2.5 GeV.
The simulated current gain is 129 $\Omega$ with RI = 360 $\Omega$.

\subsection{Preamplifier characterization}
\subsubsection{High gain version}
\begin{figure}[htbp]
\centerline{\includegraphics*[trim= 0 340 0 340,scale=0.7]{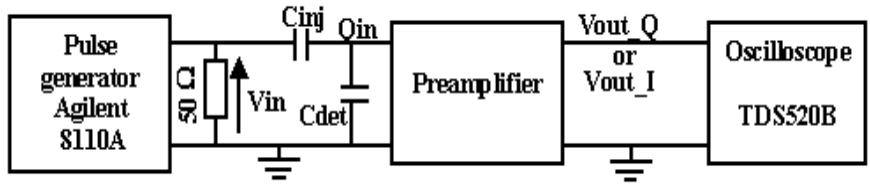}}
\caption{Injection circuit used for the measurement of gain and bandwidth.}
\label{Fig3}
\end{figure}

\begin{figure}[htbp]
\centerline{\includegraphics*[trim= 0 340 0 340,scale=0.7]{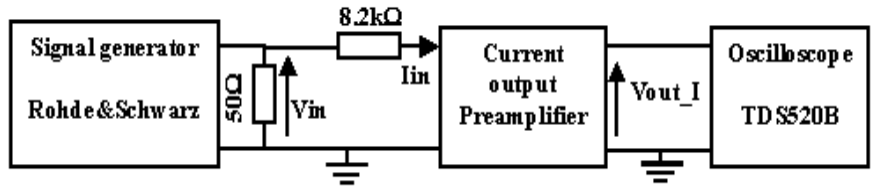}}
\caption{Injection circuit used for gain measurement.}
\label{Fig4}
\end{figure}

For the charge output the injection mode detailed in figure~\ref{Fig3} has
been used. The measured charge input is 536 mV for a current injection
equivalent to 40 MeV. This corresponds to a gain of 13.4 mV/MeV which is in
good agreement with simulation results (see 3.1.2). The decay time constant
is 33 $\mu$s (Rf = 10 M$\Omega$ and Cf = 33 pF). Figure 5 shows signals (in
relation with the current output) obtained with the injection mode of 
figure~\ref{Fig4}.

\begin{figure}[htbp]
\centerline{\includegraphics*[scale=0.8]{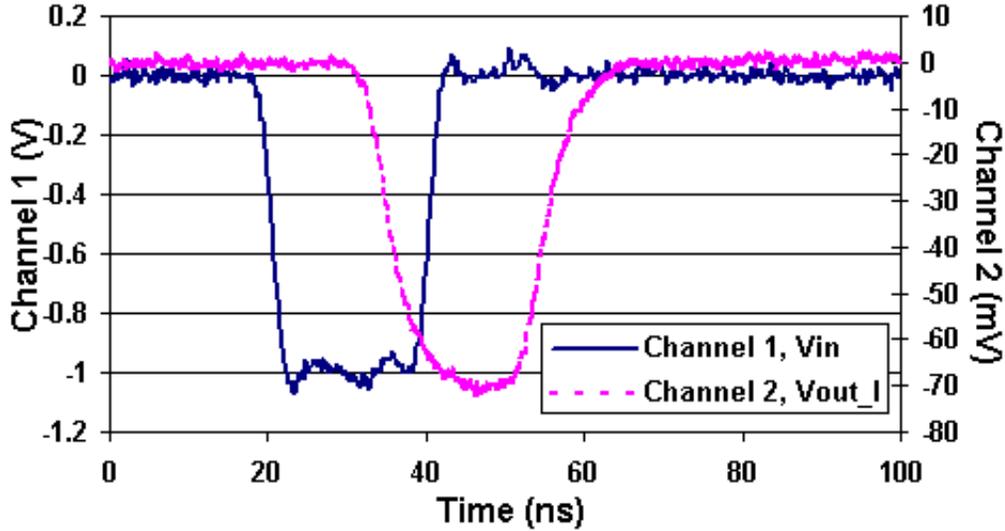}}
\caption{The injection mode is illustrated in figure~\ref{Fig4}. Channel 1
 corresponds to voltage
injection (Vin in figure~\ref{Fig4}) and channel 2 to "current output"
of the high gain preamplifier
(Vout\_I, figure~\ref{Fig1}).}
\label{Fig5}
\end{figure}

\begin{figure}[htbp]
\centerline{\includegraphics*[scale=0.8]{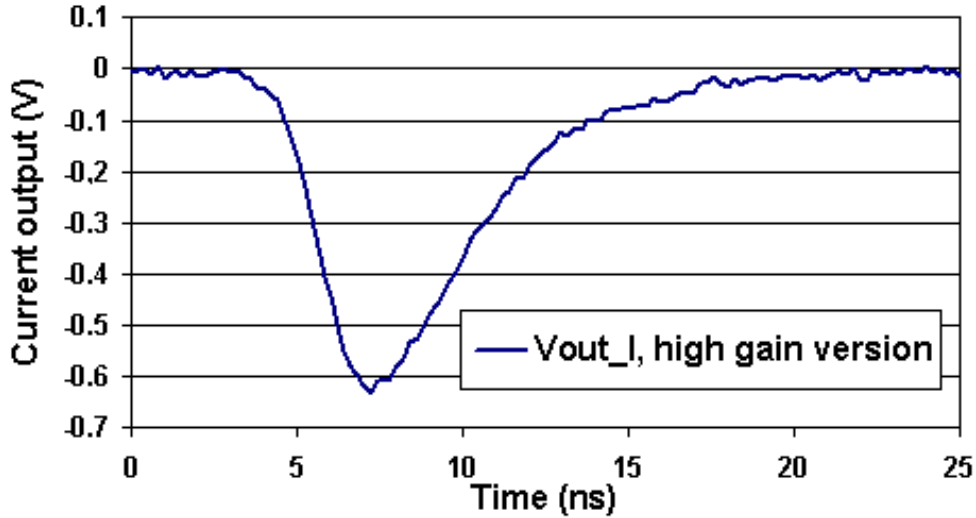}}
\caption{Current output (Vout\_I, figure~\ref{Fig1}) of the preamplifier
for the injection mode of figure~\ref{Fig3} but with $C_{det}=0$.}
\label{Fig6}
\end{figure}

Channel 1 shows the injection signal Vin. The preamplifier input
impedance being much lower than the series resistance (8.2 k$\Omega )$, the
input current (Iin in figure~\ref{Fig4}) can be easily determined.
 Channel 2
(figure~\ref{Fig5}) shows the signal obtained at the ``current'' output of the
preamplifier (Vout\_I in figure~\ref{Fig1}). The measured gain is then 574 V/A.
Concerning the time response of the preamplifier,
figure~\ref{Fig6} shows the signal obtained at the current
output (diagram of
figure~\ref{Fig1}) when the circuit of figure~\ref{Fig3} is used without Cdet
(capacitance used to simulate the detector). In order to be close to
a $\delta$ - impulse, a high speed pulse generator has been used.
 Figure~\ref{Fig6} shows the current
output signal: a 2 ns rise time is observed.
\subsubsection{Low gain version}
\begin{figure}[htbp]
\centerline{\includegraphics*[scale=0.8]{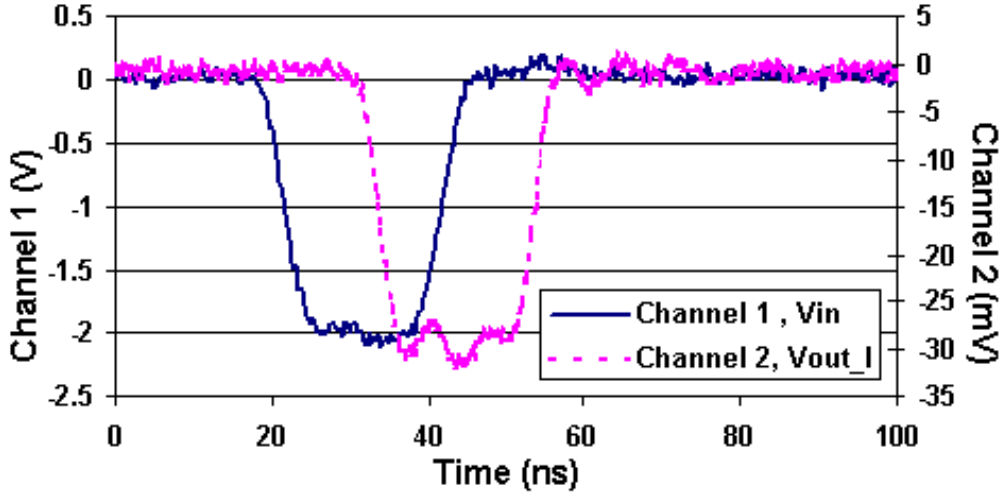}}

\caption{The injection mode is illustrated in figure~\ref{Fig4}. Channel 1
  shows voltage injection
(Vin in figure~\ref{Fig4}), channel 2 the "current output" of the
preamplifier
(Vout\_I, figure~\ref{Fig2}).}
\label{Fig7}
\end{figure}

Injection circuits are the same as for subsection 3.2.1.
For the charge output the measured
gain is $\sim$1.4 mV/MeV and the decay time constant
is 450$\mu$s; the bridge R1/R2 increases the RfCf time constant by a factor 
1+R1/R2.
For the current output
figure~\ref{Fig7} shows the observed signals.
The gain is 115 V/A. This gain is very close to the simulated gain (see 3.1.3.)
The proper preamplifier rise
time for the low gain version is also 2 ns.
\section{Beam tests}  
\subsection{Experimental details}
Beam tests have been performed at the TANDEM accelerator of the
Institut de Physique Nucl\'eaire in Orsay (France).
Protons and deuterons with energies
3 and 5 MeV and carbon isotopes ($^{12}C$ and $^{13}C$) with energy 80 MeV
were elastically scattered on a 200 $\mu$g/$cm^2$ Au
target. Two totally depleted
and overbiased n-type Si detectors in reverse mount (rear contact as
entrance window) were used. They were
manufactured by Canberra Eurisys (specifications are given in
table~\ref{table1}) and
processed by the adapted preamplifier prototypes.
The high gain (low gain) preamplifier
was coupled to the detector dedicated to structure studies (dynamics studies)
to detect particles (carbon ions).The distance between the 
detector and the preamplifier was reduced to minimum using a single connector.
The block scheme of the electronics is shown in figure~\ref{Fig10}.
Current pulse measurements were performed with a large bandwidth digital
oscilloscope (4 Gsamples/s, 8 bits) connected to the general acquisition
system via
a GPIB-VME interface; oscilloscope sensitivities were respectively 1, 2 
and 5 mV/div. for incident energies on detectors 3, 5 and 80 MeV.
A classical spectroscopy chain was used for the charge
output.
\begin{table}
\caption{Surface s, thickness x, resistivity $\rho$,
depletion bias $V_{\mathrm d}$ and operation bias $V_{\mathrm o}$ of
detectors used in our tests.}  
\begin{center} 
\begin{tabular}{ccccccc}
\hline Dedicated to & silicon type & s [mm$^2$] & x [$\mu$m] &
$\rho$[K$\Omega$cm] & $V_{\mathrm d}$[V] & $V_{\mathrm o}$[V] \\ 
\hline 
structure studies & normal & 64 & 300 & 12.7 & 28 & 90  \\ 
dynamics studies& n-TD & 200 & 300 & 2.5 & 128 & 190 \\ 
\hline 
\end{tabular} 
\end{center}
\label{table1}
\end{table}
\begin{figure}[htbp]
\centerline{\includegraphics*[trim= 0 320 0
320,scale=0.9]{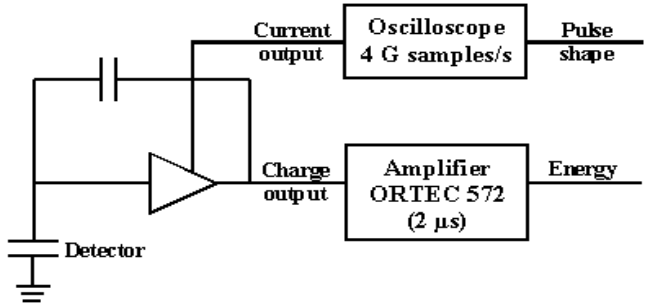}}
\caption{The electronics used in the experiment}
\label{Fig8}
\end{figure}
\subsection{Results, simulations and discussion}
First of all energy resolutions were measured from charge outputs using
scattered protons ($^{12}C$) with energy 5 MeV (80 MeV). Resolution values
(FWHM) are 30 keV and 0.3\% for respectively protons and $^{12}C$, which fully
comply with the requirements (see section 2). 
By comparison with charge outputs, energy values obtained by integrating
current signals and using
measured current gains (see 3.2) are respectively 4.97 and 4.92 MeV for
incident 5.00 MeV protons and deuterons and 79.3 MeV for incident 80 MeV 
$^{12}C$
ions; associated measured energy resolutions are in the range 2.5-3.5\%.
\begin{figure}[htbp]
\includegraphics*[width=7.cm]{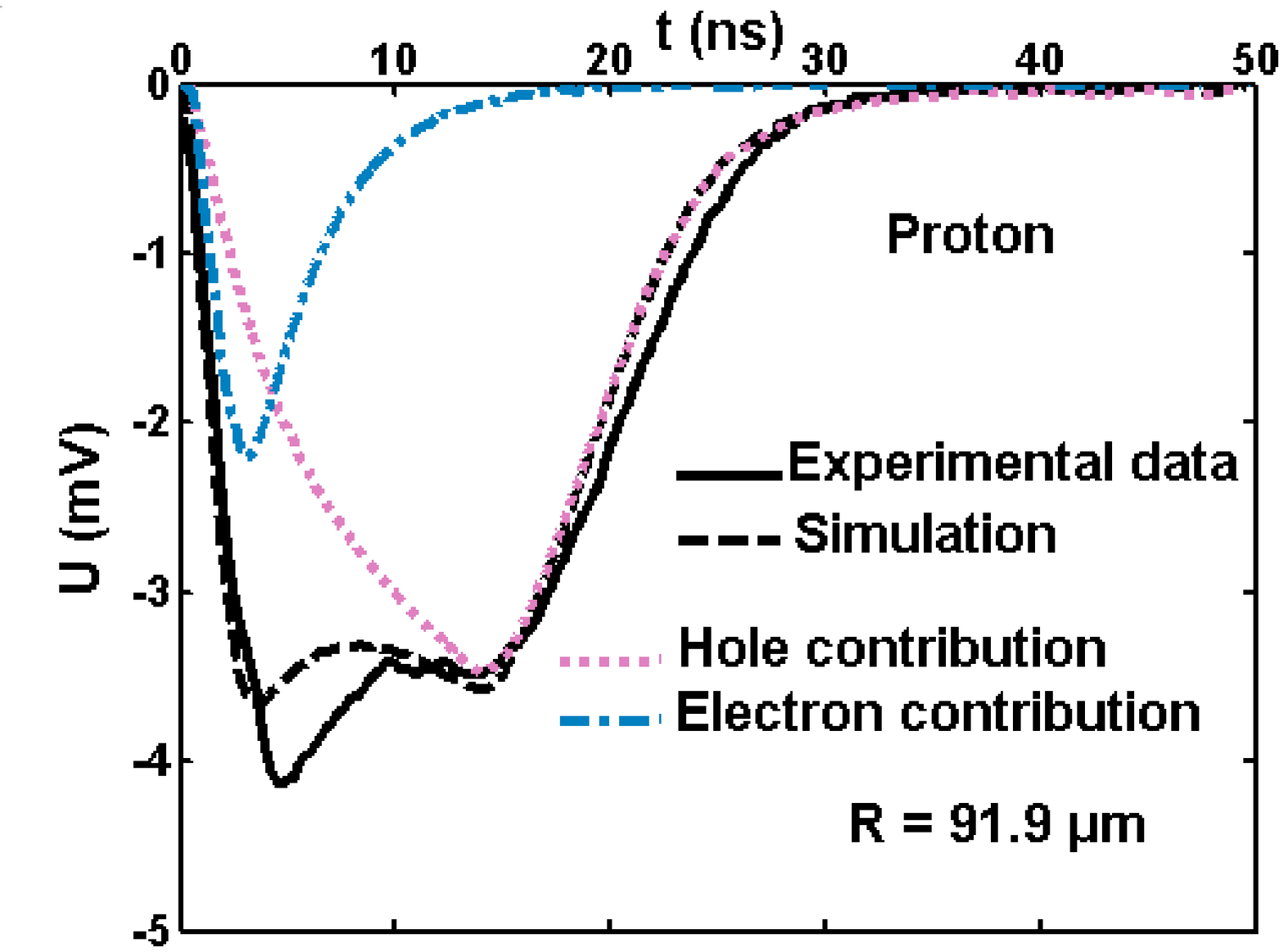}
\includegraphics*[width=7.cm]{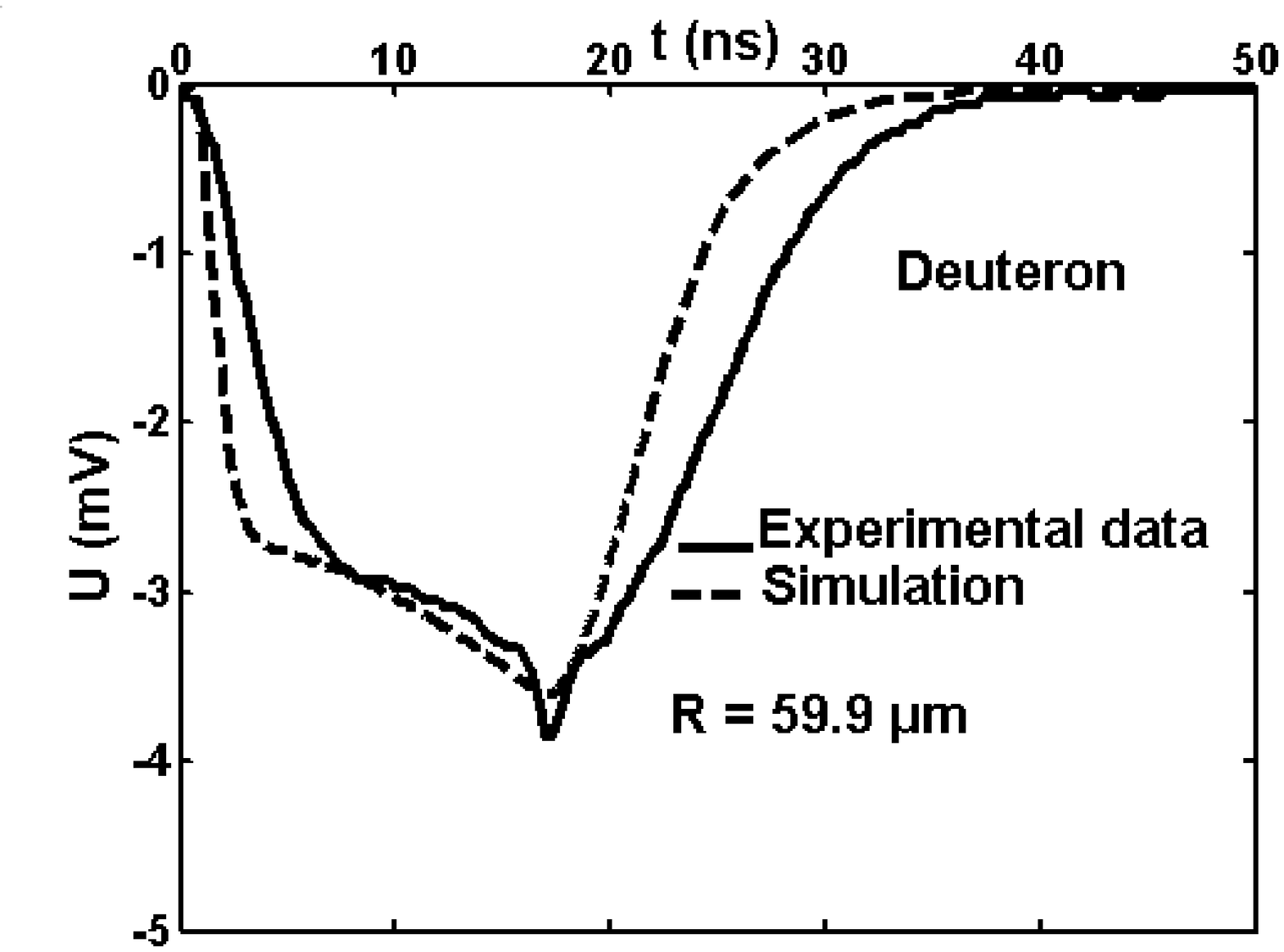}
\caption{Mean experimental current signals and simulations for (left)
  3 MeV protons
  and (right) 3 MeV deuterons. R indicates the range of the considered
particles (from~\cite{ZIG85}).}
\label{Fig9}
\end{figure}
Figures~\ref{Fig9} and \ref{Fig10} show experimental signals obtained
at the current preamplifier (high gain) output 
for 3 and 5 MeV protons and deuterons. The displayed signals correspond to
the average of about a thousand pulses.
At 3 MeV, in particular, we can notice
the peculiar shape of signals. We shall come back to this point later.
Those results show clearly at 5 MeV that for two different impinging particles
(proton and deuteron) with the same energy, the
 signals  from the detector have the
same duration but different shapes which can be used for discrimination.
With a standard preamplifier
(charge output) which integrates current signals, we
would not be able to easily discriminate both particles because of the similar
rise times.
\begin{figure}[htbp]
\includegraphics*[width=7.cm]{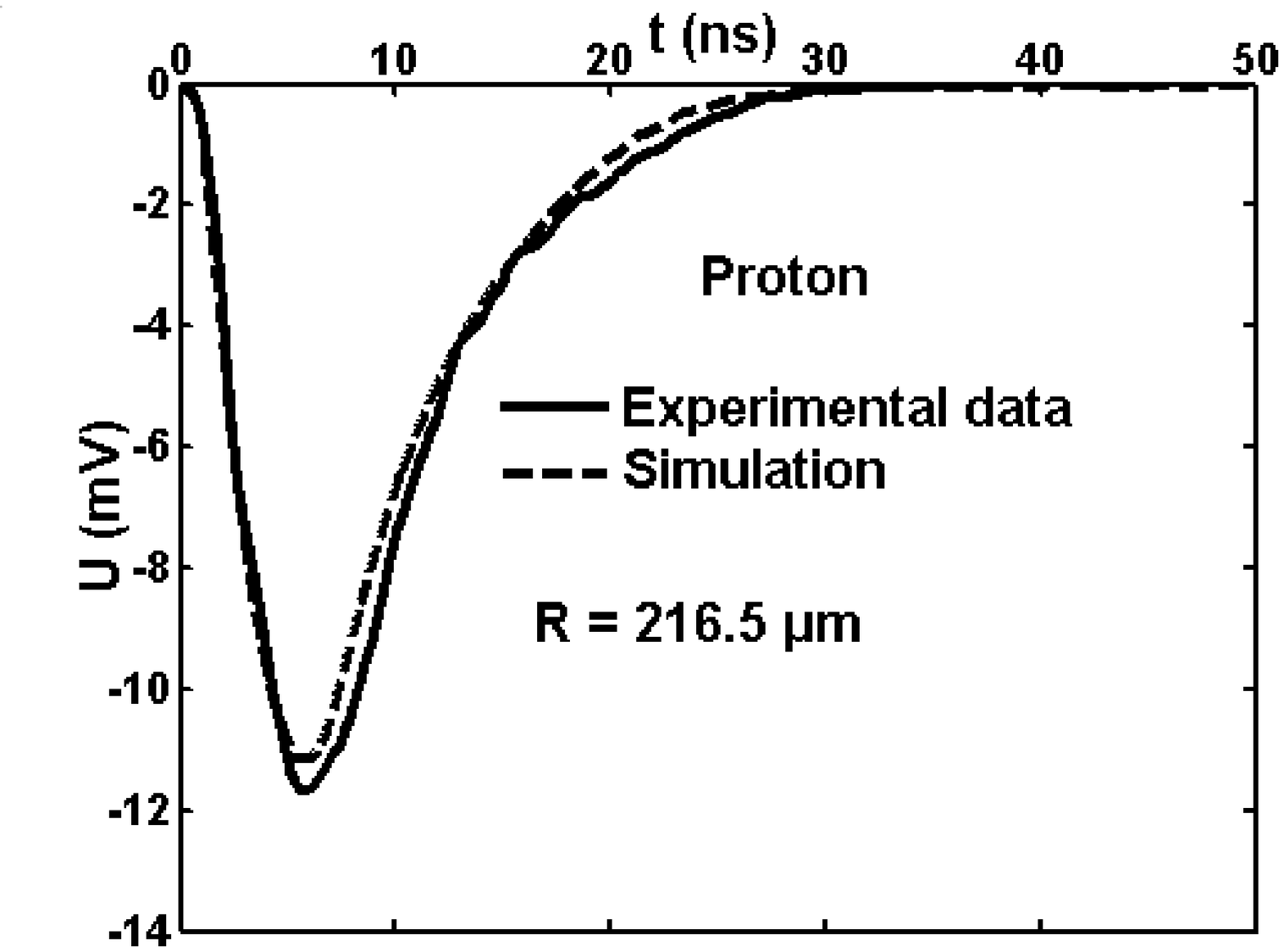}
\includegraphics*[width=7.cm]{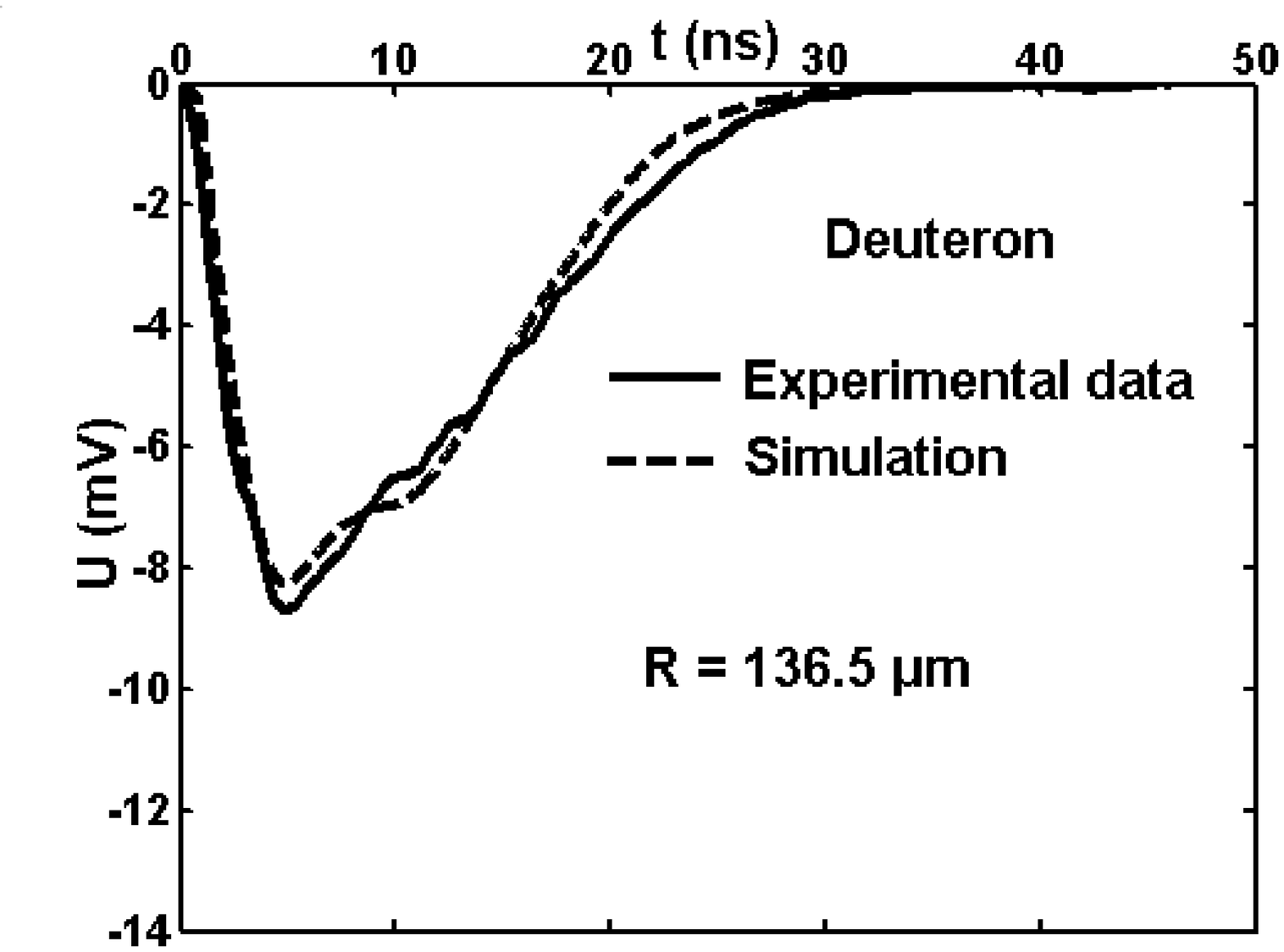}
\caption{Mean experimental current signals and simulation for (left) 5 MeV protons
  and (right) 5 MeV deuterons. R indicates the range of the considered
particles (from~\cite{ZIG85}).}
\label{Fig10}
\end{figure}

A simple simulation is compared to the signals.
To produce the output current pulse, we need two parts in simulations : one
giving the current pulse delivered by the detector and one which simulates
the preamplifier. The first one consists in describing the
drift, in the electric field, of charge carriers produced all along the
track of the particle. Normalized stopping powers for particles
(corresponding to energy losses in bins of elementary thickness
$\Delta$x= particle range/200) are calculated using 
SRIM~\cite{ZIG85}. An electon-hole pair is created by 3.6 eV energy deposit
and the mobility for holes and electrons as a function
of the electric field is calculated using the empirical equation of 
reference~\cite{LE99}. Then the current output signal from the preamplifier
(homogeneous to a voltage) is obtained calculating the preamplifier
response to the time dependent signal from the detector; 
AWBHDL software developed by CADENCE ELECTRONICS is used.
For this second part of the simulation we also need the
impedance of the detector seen by the preamplifier. Table~\ref{table2} shows
the values measured for the different elements corresponding to
the two detectors; measurements were done using a vector impedance
meter (HP 4815 RF) at different frequencies (1, 10, 30, 70 MHz).
Results of simulations are compared 
\begin{table}
\caption{Resistance R, capacitance C and inductance L of detectors}  
\begin{center} 
\begin{tabular}{cccc}
\hline Dedicated to & R [$\Omega$] & C [pF] & L [nH]\\
\hline 
structure studies & 2 & 16 & 3  \\ 
dynamics studies& 4 & 62 & 3 \\ 
\hline 
\end{tabular} 
\end{center}
\label{table2}
\end{table}
to the experimental current signals obtained for 3 (figure~\ref{Fig9}) and
5 MeV (figure~\ref{Fig10}) protons and deuterons. The calculated pulses 
give a good account of the measured signals.
\begin{figure}[htbp]
\centerline{\includegraphics*[scale=0.85]{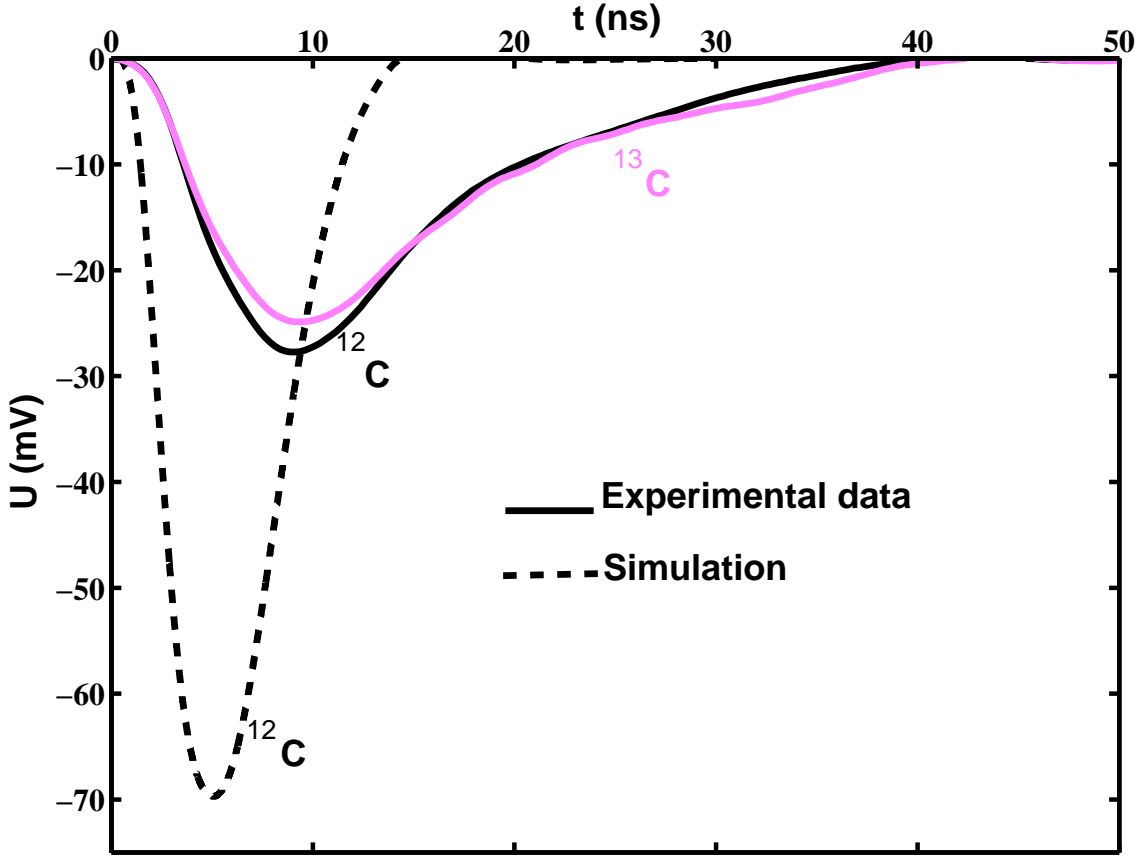}}
\caption{Mean experimental current signals for 80 MeV $^{12}C$ and
 $^{13}C$; simulation for $^{12}C$ is also presented.}
\label{Fig11}
\end{figure}
\begin{figure}[htbp]
\centerline{\includegraphics*[scale=1]{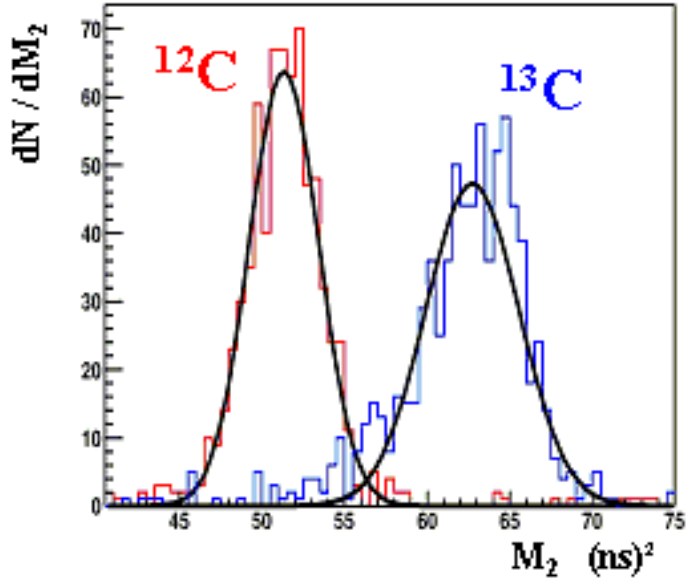}}
\caption{Discrimination of carbon isotopes. Histograms correspond to the
distributions of the second moment values $M_2$ of pulses and full lines to
gaussian fits.}
\label{Fig12}
\end{figure}
Depending on the range of
particles, contributions to signals from hole and electron drifts are more
or less visible. The smaller is the range the larger is the difference in time
between contributions to the signal from electrons (faster contribution) and
holes. An example of the simulated contributions from electrons and holes to
the current is displayed for protons in figure~\ref{Fig9}.

For 80 MeV carbon isotopes, figure~\ref{Fig11} displays the observed
mean current signals (averaged over about thousand recorded pulses). Once again pulses corresponding to the two isotopes have
the same duration but their shapes are slightly different and we can
notice in particular the variation of amplitude. An example of identification
of the two isotopes using the second moment of current pulses is shown in
figure~\ref{Fig12}. Results of the simple simulation are also given in
figure~\ref{Fig11} and a big disagreement is observed. This observation is
not surprising. Indeed it is well known that
for heavy ions the density of ionization is much higher and consequently
the extraction of electrons and holes from the zone of high carrier density
is delayed and reduced due to a screened electric field (see for
example~\cite{PA94} and references therein). Simulations 
taking those facts into account
 in a fully consistent way are in progress~\cite{PA03}. The extension of
identification possibilities to heavier and more energetic ions is currently
under study.
\section{Conclusions} 
Charge and current-sensitive preamplifiers well suited for pulse shape
discrimination techniques have been developed and successfully tested.
The feasiblity of identification from current pulse shapes for light charged
particles is clearly established. Moreover it was shown that a simple
simulation, which has to be
improved (3D simulation including resistivity variations) can help in the
design definition for future detectors. For heavy ions the work presented here
is a starting point for more complete investigations including both the
constitution of a data base of current signals (different ions at various
incident energies) and a better modeling ( complete description in terms
of finite elements) of the charge collection in presence of a high
ionization density. The obtained results also suggest the replacement of
analog methods with digital sampling techniques~\cite{BA02}. 
\section{Acknowledgements} 
The authors are indebted to CNES-DESER and ONERA-DESP for providing us with 
their scattering chamber at the TANDEM accelerator.


\end{document}